# Electronic structure, charge transfer, and intrinsic luminescence of gadolinium oxide nanoparticles: Experiment and theory

D. A. Zatsepin[1,2], D. W. Boukhvalov[3,4], A. F. Zatsepin[1], Yu. A. Kuznetsova[1], M. A. Mashkovtsev[1], V. N. Rychkov[1], V. Ya. Shur[5], A. A. Esin[5], E. Z. Kurmaev[1,2]

[1]*Institute of Physics and Technology, Ural Federal University, Mira Str. 19, 620002 Yekaterinburg, Russia*
[2]*M.N. Miheev Institute of Metal Physics of the Ural Branch of the Russian Academy of Sciences, 18 Kovalevskoj Str., 620990 Yekaterinburg, Russia*
[3]*Department of Chemistry, Hanyang University, 17 Haengdang-dong, Seongdong-gu, Seoul 133-791, Korea*
[4]*Theoretical Physics and Applied Mathematics Department, Ural Federal University, Mira Street 19, 620002 Yekaterinburg, Russia*
[5]*Institute of Natural Sciences, Ural Federal University, 51 Lenin Ave, 620000 Yekaterinburg, Russia*

**Abstract**

The cubic (c) and monoclinic (m) polymorphs of $Gd_2O_3$ were studied using the combined analysis of several materials science techniques – X-ray diffraction (XRD), scanning electron microscopy (SEM), X-ray photoelectron spectroscopy (XPS), and photoluminescence (PL) spectroscopy. Density functional theory (DFT) based calculations for the samples under study were performed as well. The cubic phase of gadolinium oxide (c-$Gd_2O_3$) synthesized using a precipitation method exhibits spheroidal-like nanoclusters with well-defined edges assembled from primary nanoparticles with an average size of 50 nm, whereas the monoclinic phase of gadolinium oxide (m-$Gd_2O_3$) deposited using explosive pyrolysis has a denser structure compared with natural gadolinia. This phase also has a structure composed of three-dimensional complex agglomerates without clear-edged boundaries that are ~21 nm in size plus a cubic phase admixture of only 2 at. % composed of primary edge-boundary nanoparticles ~15 nm in size. These atomic features appear in the electronic structure as different defects ([Gd…O–OH] and [Gd…O–O]) and have dissimilar contributions to the *charge-transfer* processes among the appropriate electronic states with ambiguous contributions in the Gd 5p – O 2s core-like levels in the valence band structures. The origin of [Gd…O–OH] defects found by XPS was well-supported by PL analysis. The electronic and atomic structures of the synthesized gadolinias calculated using DFT were compared and discussed on the basis of the well-known joint OKT–van der Laan model, and good agreement was established.

# 1. Introduction

Oxide compounds of 4f elements (lanthanides) have been attracting interest for several applications since the introduction of civil nuclear fuel technologies. In particular, $Gd_2O_3$ was used first as a stabilizing additive for the cubic phases of some neutron absorbers (or neutron poisons) in reactor fuel-cycles [1]. At that time, substantial attention was concentrated on radiation-resistant performance and the radiation-stimulated processes of decomposition within the given fuel-cycle, with a focus on the yield of radiation-inert functionalized materials.

In reality, gadolinium was never found as a pure free metal in nature up to present, so the most common chemical form the gadolinium oxide ($Gd^{3+}$) is considered. Currently, this oxide is effectively employed in other equally important applications, *e.g.*, magnetic resonance and fluorescence imaging [2], fabrication of the base for Judd-Ofelt high-efficiency luminescent devices [3], doping-modification of thermally treated nanocomposites [4], semi-commercial manufacturing of magnetocaloric materials [5-6], etc. These abundant and promising applications of $Gd_2O_3$ appear to be directly linked with its high chemical activity and structural specificity. Gadolinium oxide already has three proven structural polymorphs that had been successfully synthesized earlier [7] – the temperature-sensitive cubic phase (c-$Gd_2O_3$, with two dissimilar gadolinium lattice sites and some variety in the surrounding geometries for oxygen ligands in the natural form), the more temperature-stable monoclinic phase (m-$Gd_2O_3$, which can be fabricated *via* the cubic-to-monoclinic temperature phase-junction above 1200 ºC), and the high-temperature hexagonal phase (h-$Gd_2O_3$ [8-9], with a melting point over 2500 ºC, which is important for flame-combustion chamber manufacturing). It should be noted that the melting temperature can be essentially increased for m-$Gd_2O_3$ by Er- or/and Yb-doping with the use of ß-alanine [10]. Finally, these three $Gd_2O_3$ polymorphs exhibit a clear increase in the dielectric constant *k* in the sequence c-$Gd_2O_3$→ m-$Gd_2O_3$→ h-$Gd_2O_3$ from 14 up to 24, what is in nearly perfect coincidence with the Clausius–Mossotti molar volume versus polarizability relationship for different crystal structures [11].

XPS study of the electronic structure of gadolinium oxides and $Gd_2O_3$-based derivatives with different morphologies was originally initiated by van der Laan and co-workers [12] and then was continued by Gupta et al. [13] (both studies agree well with the later reports of XPS data in Refs. [14-15]) and Chang et al. [16]. The results of those studies suggested that the different crystal structures of the $Gd_2O_3$ polymorphs might contribute to the essentially different energy band-gaps but simultaneously with the opposite trend concerning the *k*-value, i.e., the h-$Gd_2O_3$ band-gap was determined to be approximately 5.4 eV, which is lower than that of c-$Gd_2O_3$ [17]. At the same time, the h-polymorph has the highest *k*-value in the sequence c-$Gd_2O_3 \rightarrow$ m-$Gd_2O_3 \rightarrow$ h-$Gd_2O_3$. Unfortunately, most of the works cited above did not address the valence band and valence base-band electronic structure areas except that reported for h-$Gd_2O_3$ [16]. Undoubtedly, these electronic structure areas are of major significance for altering the band-gap in relation to engineering the electronic structure of functionalized materials. From the experimental data accumulated to date, it is known that the Gd 4f and 5p electronic states are localized in the vicinity of 8 eV and 21 eV for the binding energies (BE) for the pure metal [14-15], respectively. It is then reasonable to expect the appropriate Gd 4f (~8 eV) ↔ O 2p (~7 eV) and Gd 5p (~21 eV) ↔ O 2s (~21.3 eV) states coupling with their BE shifts for $Gd_2O_3$ polymorphs in the final electronic structure. Therefore, one might assume that the coupling and alignments of these initial bands will depend essentially on the atomic structure not only because of the synthesis method, but also due to the lifetime broadening of the photoionized electronic shells with the subsequent multi-interaction decays via the dominating super Coster-Kronig (s-CK) channel of the 4d – 4f4f type ($N_{4,5} - N_{6,7}N_{6,7}$) in the excited electronic states [12]. Thus, in this specific situation, the variety of final electronic band alignments, charge states of the gadolinium central atom and structural geometry of the surrounding oxygen ligands will be some of the most fundamental physical points for characterizing the electronic structure of a material under study as a host matrix for later technological treatments.

In the current paper, we present a study of the atomic and electronic structures of $Gd_2O_3$ polymorphs (cubic and monoclinic) synthesized by combustion-explosive microwave pyrolysis. This technological approach appears to be a promising technique because of the high purity of the yielding product compared to those of conventional chemical synthesis methods. The final c- and m-$Gd_2O_3$ polymorphs will be characterized experimentally with the help of X-ray diffraction (XRD), scanning electron microscopy (SEM), X-ray photoelectron spectroscopy (XPS), and photoluminescence (PL). The partial and total densities of states (PDOS and TDOS) will be calculated, and the final electronic structure will be modeled using a density functional theory (DFT) approach in order to obtain the most objective version of the electronic structure. Particular attention will be paid to the origin of defects in the structure of the synthesized gadolinium oxides.

## 2. Sample preparation, experimental and calculation details

Gd metal of 99.6% purity (Ural Federal University, Yekaterinburg, Russia) was used as the initial component for the fabrication of two conforming gadolinium oxide phases: cubic and monoclinic. The cubic $Gd_2O_3$ phase (denoted further as c-$Gd_2O_3$) was precipitated from the appropriate hydroxide on a clean and dry noble metal wafer (technological temporary base). Before starting the precipitation process, 200 ml of de-ionized and distilled water was swamped into the chemical reactor, where ***pH*** = 8.5 had been permanently supported and was retained until completion of the precipitation process. As a result, the hydroxide suspension was obtained and then immediately filtered several times using a vacuum filtering machine. Then, a drying cycle at 100 °C for 12 hours was applied, followed by tempering at 1000 °C for 2 hours and then fabricating c-$Gd_2O_3$ pressed pellets (1.2 cm in diameter and 3 mm in thickness). The established crystallite size was no more than 50 nm on average.

The monoclinic $Gd_2O_3$ phase (m-$Gd_2O_3$) was obtained using an already established flame-combustion technique with glycine [18]. We dissolved the appropriate melt of the initial components in de-ionized water with an additive of 80% γ-amino-acetic acid with simultaneous careful interfusion. Then, the prepared solution was set into the chamber of a microwave reactor for drying and further initiation of the explosive pyrolysis process. Usually under these conditions, explosive pyrolysis proceeds with a large and instantaneous heat release. A fine powder fraction was obtained, from which pressed pellets of m-$Gd_2O_3$ were made with the same dimensions as reported above for c-$Gd_2O_3$. No additional thermal tempering was performed on these pellets.

Initial atomic structure measurements of the samples under study were made using an XPert Pro MPD X-ray diffractometer with a solid-state Pixel detector, primary Cu *K*α X-ray radiation source, β-filtering of the secondary X-ray beam and the full-fledged Rietveld XRD analysis by applying the XPert High-Score Plus™ software. Additionally, the scanning electron microscopy (SEM) images of the samples under study were collected using a Carl Zeiss AURIGA™ CrossBeam microscope employing a field-emission Schottky cathode as the electron source and the mode with an accelerating voltage of 5 kV.

The study of the electronic structure was performed employing a Thermo Scientific™ *K*-Alpha+™ XPS spectrometer in different modes: fast wide-scan chemical contamination analysis (survey XPS spectroscopy) and core-level and valence band (VB) structure XPS mapping. This spectrometer has a dual-beam flood gun, coupling a low-energy ion beam with very low energy co-axial electrons (less than 10 eV) to prevent sample charging during analysis (GB Patent 2411763), and a 180º double-focusing hemispherical energy analyzer with an energy resolution better than 0.28 eV. All measurements were carried out with an Al *K*α monochromatized X-ray source in an oil-free vacuum at $5 \times 10^{-6}$ Pa pressure (provided by two 250 l/s turbo-molecular pumps) with a 300 μm X-ray spot diameter, a 200 eV pass-energy window for the analyzer in fast-scan mode (XPS survey) and 50 eV for the core-level and VB mapping modes, and multi-point spectra acquisition (128-channel detector) followed by summation and averaging. The samples were softly sputtered in vacuum before measurements with Ar$^+$ ions having the energy of 200 eV, 30° angle to samples surface and 10 seconds sputtering time.

As an additional and independent experimental research stage, the luminescence properties of m-Gd$_2$O$_3$ and c-Gd$_2$O$_3$ were studied. Luminescence excitation and luminescence spectra were recorded using a Perkin Elmer™ LS 55 spectrophotometer at room temperature. The spectral widths of the input and output monochromator slits were set to 2.5 nm, and the secondary electron multiplier voltage was 650 V. This measurement mode allows to obtain the signal-to-noise ratio better than 1000:1. The more detailed specifications of all employed experimental equipment might be found at the appropriate manufacturer's web-sites.

Density functional theory (DFT) calculations were performed using the SIESTA pseudo potential code [19]. All calculations were performed with the Perdew-Burke-Ernzerhof variant of the generalized gradient approximation (GGA-PBE) [20] for the case of a slab and exchange-correlation potential accounting for the dipole correction [21]. The ground electronic state was consistently found during optimization using norm-conserving pseudopotentials [22] for the cores and a double-$\xi$ plus polarization basis of localized orbitals for Gd and O. Then, all atomic positions and lattice constants were optimized. The forces and total energies were optimized as well, with accuracies of 0.04 eV/Å and 1.0 meV, respectively. For modeling of the pristine and oxygen deficient cubic phase, we used a supercell of 40 atoms (Gd$_{16}$O$_{24}$, Fig. 1a), and for modeling of the surface, we used a slab of the same supercell doubled along the c-axis (the thickness of the slab was approximately 9.5 nm, see Fig. 1b). For modeling of the bulk of the monoclinic phase, we employed a supercell consisting of 15 atoms (Gd$_6$O$_9$), and for modeling of the surface, we used a slab of the same supercell with a thickness of approximately 0.86 nm. Then, the metallic gadolinium contribution was evaluated; we performed calculations for a 6 layer slab of the *fcc* phase of Gd passivated by hydroxyl (-OH) groups. All calculations were carried out with an energy mesh cut-off of 300 Ry and a *k*-point mesh of 6×6×4 in the

Monkhorst-Pack scheme for the *bulk* and 6×6×2 for the *surface* [23]. In order to plot the final densities of states (DOS), the *k*-point mesh was increased to 8×8×6 and 8×8×4, respectively.

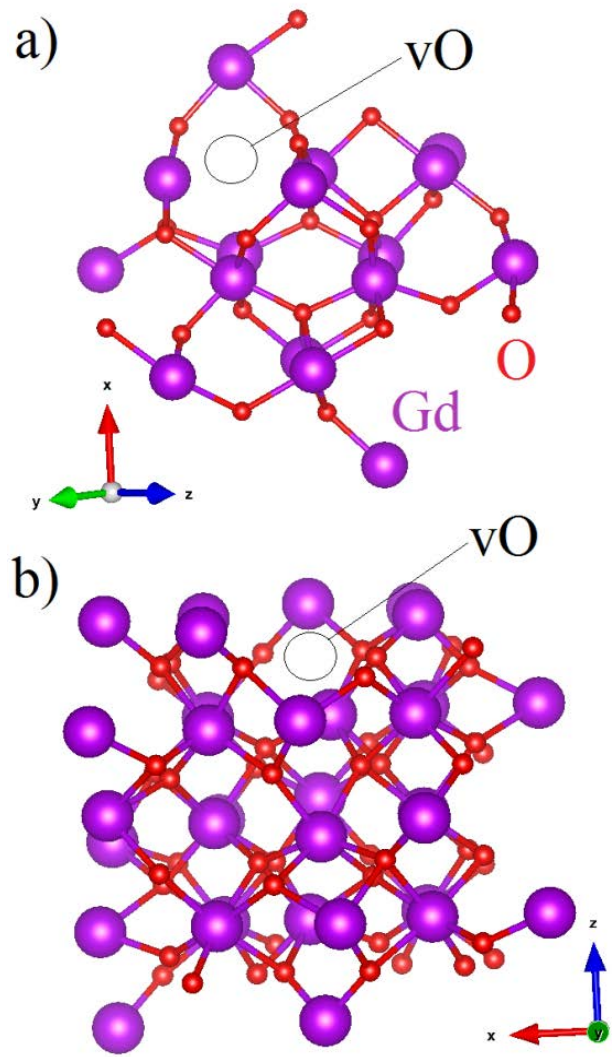

**Figure 1.** An optimized atomic structure of supercells for the bulk (a) and surface (b) cubic $Gd_2O_3$ with oxygen vacancies (vO).

## 3. Results and Discussion

*3.1 XRD and SEM structure characterizations*

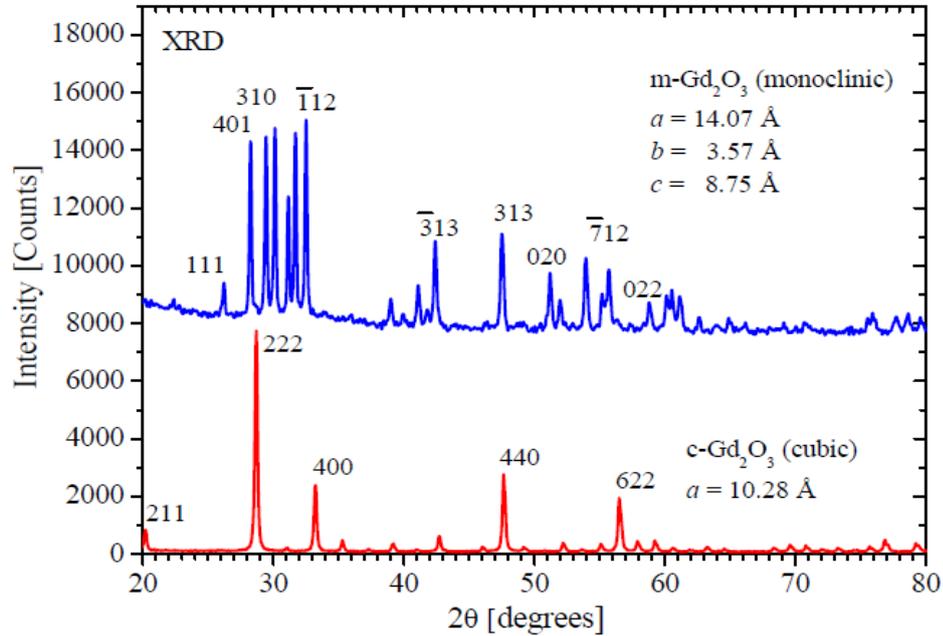

**Figure 2.** X-ray diffraction data of the cubic (c-Gd$_2$O$_3$) and monoclinic (m-Gd$_2$O$_3$) phases of the just synthesized gadolinia samples.

X-ray diffraction data (XRD) obtained for the samples under study are shown in Fig. 2. From the XRD analysis it is following, the Gd$_2$O$_3$ sample synthesized by the precipitation technique displays diffraction patterns specific to the cubic phase of gadolinia without admixtures of any other phases (Fig. 2, see red XRD pattern). Therefore, this sample was identified as c-Gd$_2$O$_3$, having the unit-cell parameter $a = 10.82$ Å. The second sample have been synthesized using the explosive pyrolysis technique and displays a majority of the monoclinic phase with the lattice constants $a = 14.07$ Å, $b = 3.57$ Å, $c = 8.75$ Å and angles $\alpha = 90°$, $\beta = 100°$, $\gamma = 90°$ with an admixture of not more than 2 at. % of the defective cubic gadolinia phase as a side effect (Fig. 2, blue XRD pattern). This is not due to the employed m-Gd$_2$O$_3$ synthesis technique because it is well-known that it is not possible to exclude completely the undesirable appearance of the more energetically favorable cubic phase as an additive, even when using standard and accepted synthesis approaches. Moreover, the concentration of the undesired cubic phase is essentially higher than 2 at. % in some specific cases [24-26]. The densities the just synthesized c-Gd$_2$O$_3$ and m-Gd$_2$O$_3$ were estimated based on XRD data analysis as 7.61 and 8.31 g/cm$^3$, respectively. The obtained values are slightly less in comparison with the data indicated in the ICSD database (7.81 and 8.35 g/cm$^3$) for cubic and monoclinic phases of Gd$_2$O$_3$ (standard cards No. 00-043-1015 and No. 01-073-2652) [27]. This result is likely due to the lattice constants of just synthesized samples which a bit exceeds values indicated in the standard cards and, correspondingly, c-Gd$_2$O$_3$ and m-Gd$_2$O$_3$ under investigation have larger unit cell volumes that leads to lower density values. It should be noted, the XRD pattern of the second sample (see Fig. 2, blue XRD pattern) confirms that no signs of a third phase, such as Gd(OH)$_3$ or Gd(OH)$_x$, were established in the XRD of m-Gd$_2$O$_3$, which should be present in the synthesized structure as a signature of the employed technological process. This probably happens due to the relatively small concentration of OH$^-$ groups (below the sensitivity limit of the XRD system used here), their stochastic and disordered distribution in the atomic structure of the sample, or because of both reasons simultaneously. The established crystallite size was 21 nm on average with a 2 at. % admixture of the cubic phase with an average crystallite size of not more than 15 nm.

Previously, XPS and XRD combined analyses were successfully applied in the case of the electronic structure study of complex Gd@C$_{82}$(OH)$_x$ compounds with different concentrations of OH$^-$ groups [28] where XRD as well did not recognize quite small OH$^-$ concentrations. Therefore, an employed X-ray photoelectron spectroscopy (XPS) for complete electronic structure mapping (survey, core-levels, and valence bands) and trace element analysis will be a justified choice (the XPS element concentration sensitivity is ~ 0.3 at. %, which is sufficient for our situation and, thus, is much reliable than standard XRD analysis).

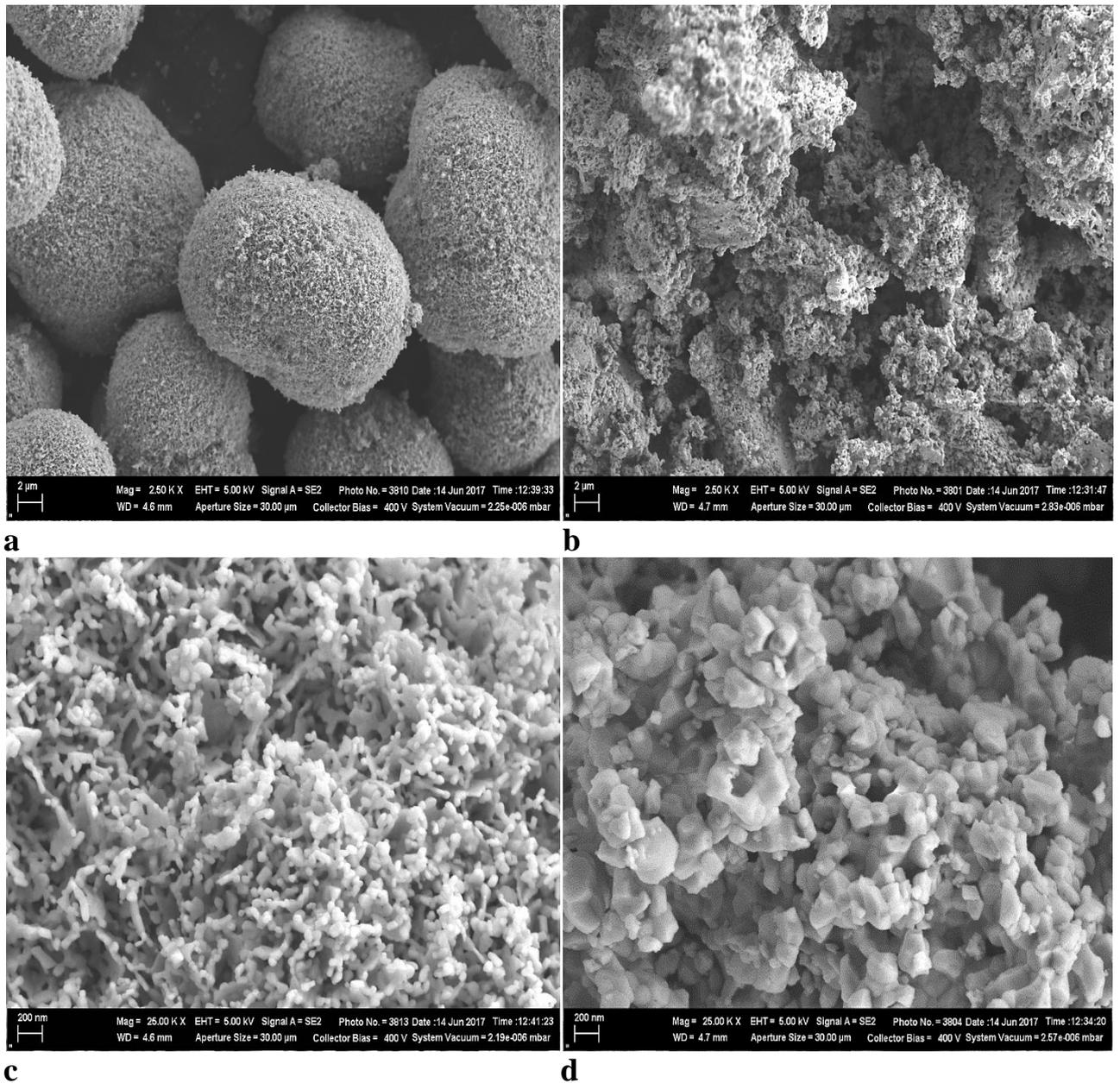

**Figure 3.** SEM images of the cubic (panels a and c) and monoclinic (panels b and d) $Gd_2O_3$ polymorphs.

Figure 3 displays the SEM images for cubic and monoclinic phases of $Gd_2O_3$ polymorphs. One can see from this figure that the cubic polymorph of the synthesized $Gd_2O_3$ is characterized by spheroidal-like agglomerates with well-defined edges and a relatively narrow distribution of nanocluster sizes on the scale of approximately 10–15 µm. These spheroidal-like nanoclusters are assembled from primary nanoparticles of the size below 200 nm. An inhomogeneous distribution of these primary nanoparticles between the surface and subsurface areas and the difficty of numerical evaluation of their size by SEM makes impossible direct quantitative comparison of the XRD and SEM data. But we can claim the good qualitative agreement between the SEM particles size evaluation (below 200 nm) and the XRD measurements (50 nm) in the samples under study. In contrast with the c-$Gd_2O_3$, the atomic assemblage of m-$Gd_2O_3$ is represented by the agglomerates with a complex three-dimensional shape without clear-edged boundaries (see Figs. 3b and 3d). In this case, the clusters are assembled from primary nanoparticles with a polyhedral shape and are estimated to have an average size of approximately 100 nm. This value is apparently overestimated if one will compare with the derivations following from the XRD analysis just reported above.

*3.2 XPS chemical analysis*

The results of the XPS survey chemical contamination analysis (wide scan) are shown in Fig. 4.

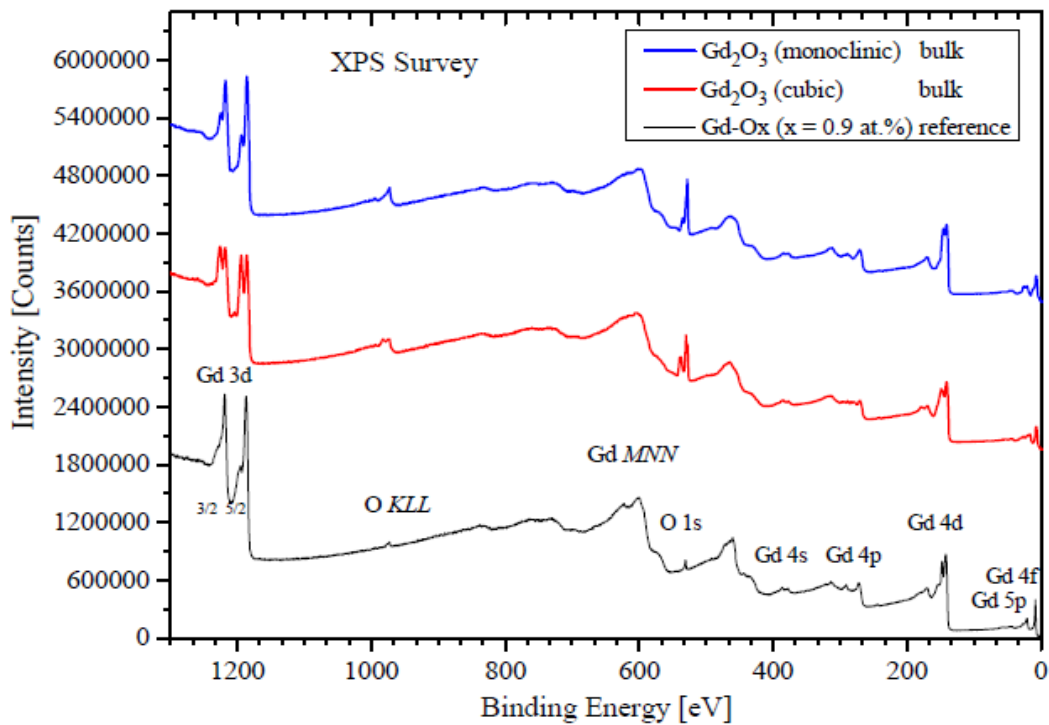

**Figure 4.** X-ray photoelectron survey spectra of m-$Gd_2O_3$, c-$Gd_2O_3$ and the Gd-$O_x$ (x = 0.9 at. %) XPS external standard.

The XPS survey spectra, core-levels, and valence bands (VB) of our samples were processed and then interpreted using the Thermo Scientific™ Avantage™ Software with additional cross-referencing of the NIST Government X-ray Photoelectron Spectroscopy Database [29], Thermo Scientific XPS Knowledge Base [30], and PDF Handbooks of Monochromatic Spectra [31]. No alien XPS signals are present in the XPS survey spectra except those belonging to the declared chemical formulas of the samples (see Fig. 4). The absence of the adventitious and vacuum hydro-carbon C 1s signal in the presented survey spectra proves the high-quality of the oil-free vacuum media that was retained during all XPS measurements. Therefore, we can declare the absence of contaminations in m-$Gd_2O_3$ and c-$Gd_2O_3$ and the validity of our samples for subsequent precise XPS electronic structure mapping.

*3.3. XPS core-level spectroscopy*

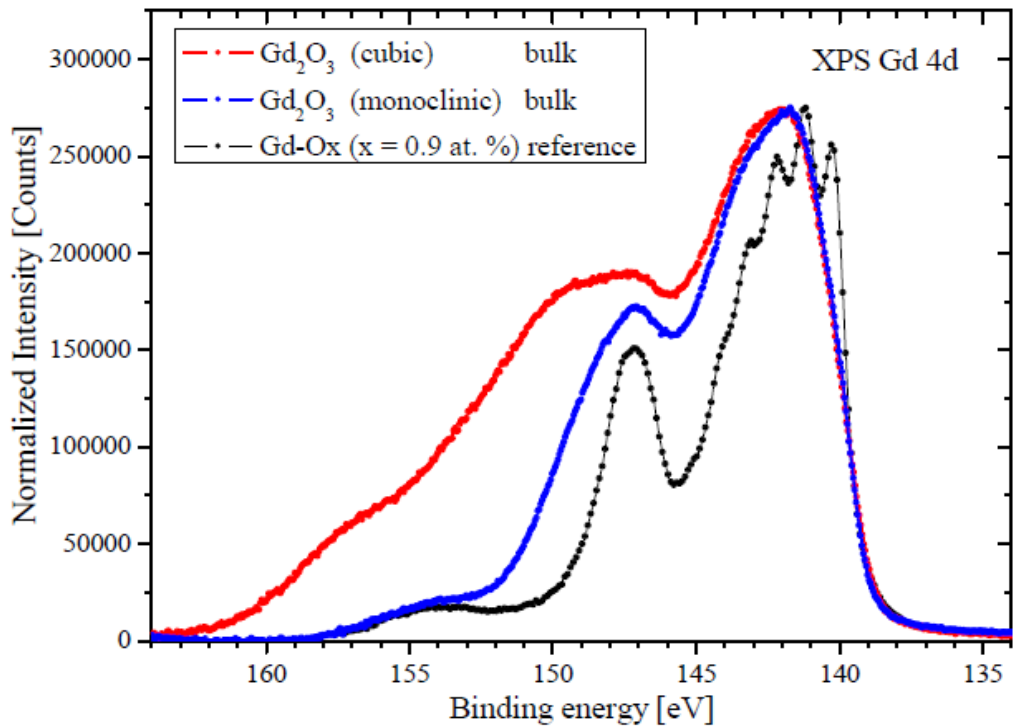

**Figure 5.** Normalized profiling of the X-ray photoelectron Gd 4d core-level spectra from c-$Gd_2O_3$ (cubic), m-$Gd_2O_3$ (monoclinic), and the Gd-$O_x$ (x = 0.9 at. %) XPS external standard.

The XPS Gd 4d core-level spectra of c-$Gd_2O_3$ (cubic, space group Ia$\bar{3}$), m-$Gd_2O_3$ (monoclinic, space group C2/m), and the slightly acidified metallic gadolinium Gd-$O_x$ (x = 0.9 at. %) XPS external standard are shown in Fig. 5. The employed Gd-$O_x$ XPS external standard clearly exhibits Gd 4d core-level line-shapes, which are nearly identical in their profiles to that of Gd metal [32] because of the very insignificant oxygen acidification (x = 0.9 at. %), whereas the synthesized m- and c-phase gadolinium oxides display essentially broader 4d core-levels. Interpretation of the obtained results was made using the momentum coupling model of van der Laan [12], which is based on the OKT approach [33]. This means that the multiplet structure clearly seen in the Gd 4d spectrum should be assigned to the strong electrostatic Gd 4d-4f interactions being predicted within the framework of the joint OKT-van de Laan model. Thus, the separation between the fine structure located in the range of 139.5 – 143.7 eV and the relatively high-intensity peak at 147.5 eV reflects the spin-orbital interactions, whereas the low-intensity and simultaneously broad band over 154 eV is linked with the *charge-transfer* processes [34]. This point agrees well with our XPS data: the dissimilar ligand geometry in c-$Gd_2O_3$ and m-$Gd_2O_3$ from the common physical consideration is the reason for the dissimilar *charge-transfer* processes in these phases. If this is true, then this *charge-transfer* in Gd-$O_x$ (x = 0.9 at. %) should be essentially weaker than in more or less completely oxidized gadolinium (*i.e.*, the ligand-deficient sample is opposite to c-$Gd_2O_3$ and m-$Gd_2O_3$). Indeed, the Gd-$O_x$ XPS Gd 4d spectrum exhibits the weakest and lowest intensity band over 154 eV (see Fig. 5), supporting our considerations in the framework of the joint OKT-van de Laan model. Additionally, different ligand surroundings of the central gadolinium atom in the c-phase and m-phase of $Gd_2O_3$ will cause dissimilar s-CK decays of the excited electronic states because of the different characteristics of the Gd-O bonds in these oxide phases. In other words, different s-CK broadening [12, 34] and distortion of these spectra will occur, as can be observed in our XPS data shown in Fig. 5 (please compare the red and blue spectra). The nearly purely metallic Gd-$O_x$ (x = 0.9 at. %, black spectrum in Fig. 5) has the lowest intensity *charge-transfer* band, supporting the assumption concerning the origin of the broad bands over 154 eV. The most narrow and well-resolved main XPS bands in the range of 139.5 – 143.7 eV and at 147.5 eV indicate the lowest s-CK distortions of the XPS line-shapes in the Gd 4d core-level spectrum.

Finally, the observed Gd 4d transformations for c-Gd$_2$O$_3$, m-Gd$_2$O$_3$, and the Gd-O$_x$ (x = 0.9 at. %) XPS external standard discussed above coincide well with the main features of the joint OKT-van de Laan model [12, 34].

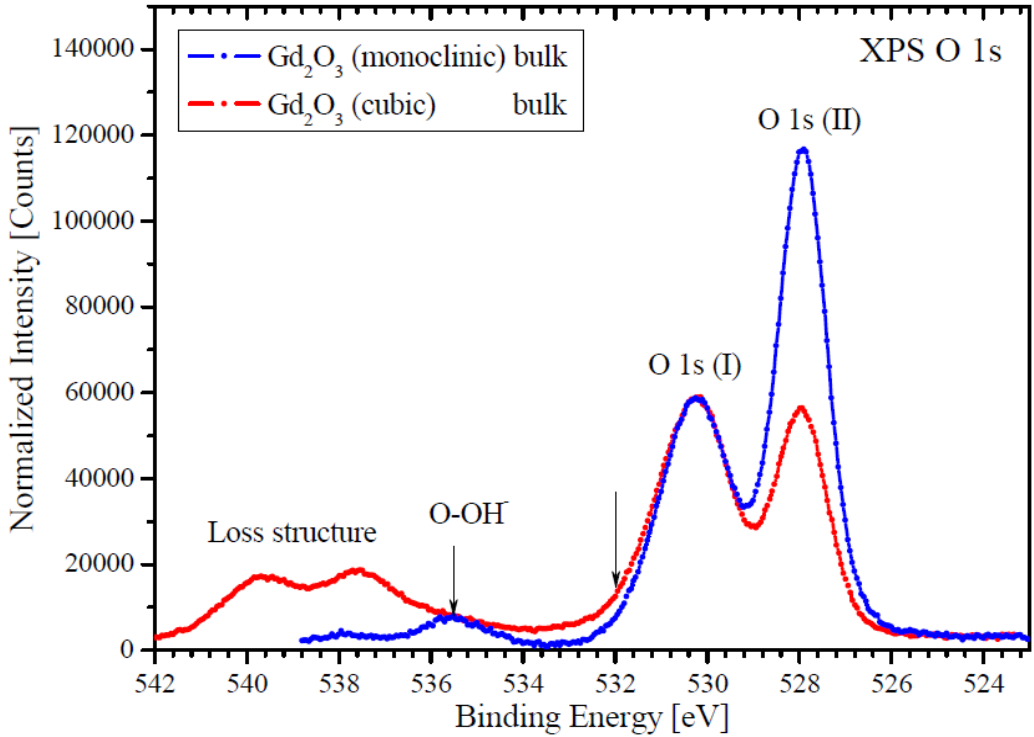

**Figure 6.** Normalized profiles of the X-ray photoelectron O 1s core-level spectra for the c-Gd$_2$O$_3$ (cubic) and m-Gd$_2$O$_3$ (monoclinic) polymorphs.

The XPS O 1s core-level spectra of the c-Gd$_2$O$_3$ (cubic) and m-Gd$_2$O$_3$ (monoclinic) polymorphs are shown in Fig. 6. In the relatively straightforward case of the oxide electronic structures, the O 1s core-levels usually exhibit fairly standard and simple single-line shapes, allowing unambiguous spectral interpretation without any additional difficulties. However, from Fig. 6 follows that we faced a complicated variant of the O 1s core-level structure. Cross-referencing with the NIST XPS Database [29], Thermo Scientific XPS Knowledge Base [30], and PDF Handbooks of Monochromatic Spectra [31] allows us to affirm with full confidence that the XPS O 1s (I) band, located at 530.2 eV, belongs to the natural (*i.e.*, lattice) oxygen-gadolinium bond. For this reason, it is selected in order to perform normalized XPS band profiling, which is a methodology first suggested by J. Kawai *et al.* [34]. In the framework of this profiling, the dissimilarity of the O 1s (I) core-level XPS bands at 532 eV (marked by the vertical arrow in Fig. 6) denotes the dissimilar characteristics of the Gd-O bonding in the cubic and monoclynic polymorphs synthesized using different methods, which is not very surprising. This deviation arises from the dissimilar technological processes employed for the fabrication of c-$Gd_2O_3$ and m-$Gd_2O_3$ and is sometimes referred to as a technological signature of the just synthesized sample. Recall that the m-phase has some admixture of the cubic phase (approximately 2 at. %) and was made with so-called "wet" technology, meaning it should contain some additives of $OH^-$ groups (see *Sample preparation...* details), which can be assigned intuitively and at first glance with the well-resolved XPS mono sub-band at 535.5 eV (see Fig. 6). The absence of this sub-band in the O 1s core-level spectrum of c-$Gd_2O_3$, which has been synthesized without employing "wet" technology, allows for this. However, the situation with OH-clustering, "metal-OH" bonding and the manifestation of this bonding in the XPS O 1s core-level spectra is not as straightforward as it seems. For the samples made by microwave synthesis, the signals in the vicinity of 535.3 eV belong to either the anomalous and isolated O–O bonding or O–OH [34]. Taking into account the "wet" synthesis procedure for the m-$Gd_2O_3$ polymorph, we are inclined to assign the 535.5 eV sub-band to [Gd…O–OH] bonding rather than O–O. If our assumption is valid, we will see the manifestation of this specific chemical bonding in the valence band spectrum, which will be shown and discussed below. The binding energy range over ~537 eV usually belongs to the XPS loss structures, and our recorded XPS feature agrees well with that had been reported in the Thermo Scientific XPS Knowledge Base [31].

The XPS band at 528.2 eV and denoted as O 1s (II) might be considered as of essential spectroscopic interest (see Fig. 6). This band is present in the O 1s core-level spectra of both the c-phase and m-phase of $Gd_2O_3$, but has dissimilar contributions to the final electronic structure of our samples. There are several satisfactory spectroscopic data concerning the structure of O 1s core-levels in gadolinium oxides; unfortunately, some of them do not agree well with each other in terms of XPS analysis. For instance, Barecca *et al.* [36] reported about the double-band O 1s core-level structure with partial contributing components located at 529.0 eV ± 1 eV ($Gd_2O_3$ lattice oxygen) and 531.5 eV ± 1 eV (technological signature of adsorbed oxygen from $OH^-$ groups and carbonates). In contrast to Barecca *et al.*, S. Majeed and S.A. Shivashankar [37] discussed the simply shaped XPS O 1s mono-band in $Gd_2O_3$, which seems questionable because of the reasons discussed above in terms of Gd-O bonding defects. Thus, we will concentrate our discussion on the XPS data of others, Barecca *et al.* in particular. Their interpretation of the O 1s core-level signal at 529.0 eV is very close to what we are observing at 530.2 eV. The 1 eV difference in binding energy positions is easily explained by the fact that Barecca *et al.* [36] employed a PHI 5600ci XPS spectrometer with a first-generation single-channel charge neutralizer and spherical-sector energy analyzer, whereas for our XPS measurements performed using a Thermo Scientific™ *K*-Alpha+™ XPS spectrometer (model year: 2016), where the BE ambiguity determination was not more than 0.2 eV (maximum deviation value). Thus, we suppose that Barecca *et al.* actually recorded a signal at 529.0 eV ± 1 eV in the O 1s core-level spectrum of $Gd_2O_3$ lattice oxygen. At the same time, we cannot agree with their interpretation of the signal at 531.5 eV ± 1 eV belonging to the adsorbed oxygen from $OH^-$ groups and carbonates because our samples do not contain carbonates or vacuum hydrocarbons (see survey spectra in Fig. 4). This interpretation is also not perfect because intrinsic oxygen defects are usually exhibiting the XPS signal in the vicinity of 532 eV. The latter was supported by our previous XPS data on $TiO_2$ and ZnO host matrices with different ion implantation modes and by an independent luminescence study of intrinsic defects in gadolinium oxides [38]. For additional support of our viewpoint on the origin of our O 1s (I) and O 1s (II) bands, the interpretation of our XPS data after Külah *et al.* [39] might be considered; their wide-and-low intensity "Gd–OH" XPS O 1s band (~533 eV), so-called "lattice" Gd–O XPS O 1s band (530 eV), and oxygen excess [39] Gd–$O_y$ (528.2 eV) are in nearly perfect agreement with our present XPS data report. In our humble opinion, the dissimilarity of spectral behavior for this oxygen excess 528.2 eV XPS band in the c-$Gd_2O_3$ and m-$Gd_2O_3$ also well agrees with the different technological synthesis method of these two gadolinia polymorphs (see *Sample Preparation...* Section) and is well supported by different type of particles packing even visually seen at SEM images (Fig. 3).

Finally, taking into account our XPS O 1s data, SEM and the results after Külah *et al.* [39], we make assumptions about at least the two types of oxygen defects in $Gd_2O_3$ – deficient and excess. This established oxygen heterogeneity is considered without accounting for the $OH^-$ group technological signatures. Our assumption will be examined below using the photoluminescence technique and DFT modeling.

*3.4. XPS Valence Band mapping*

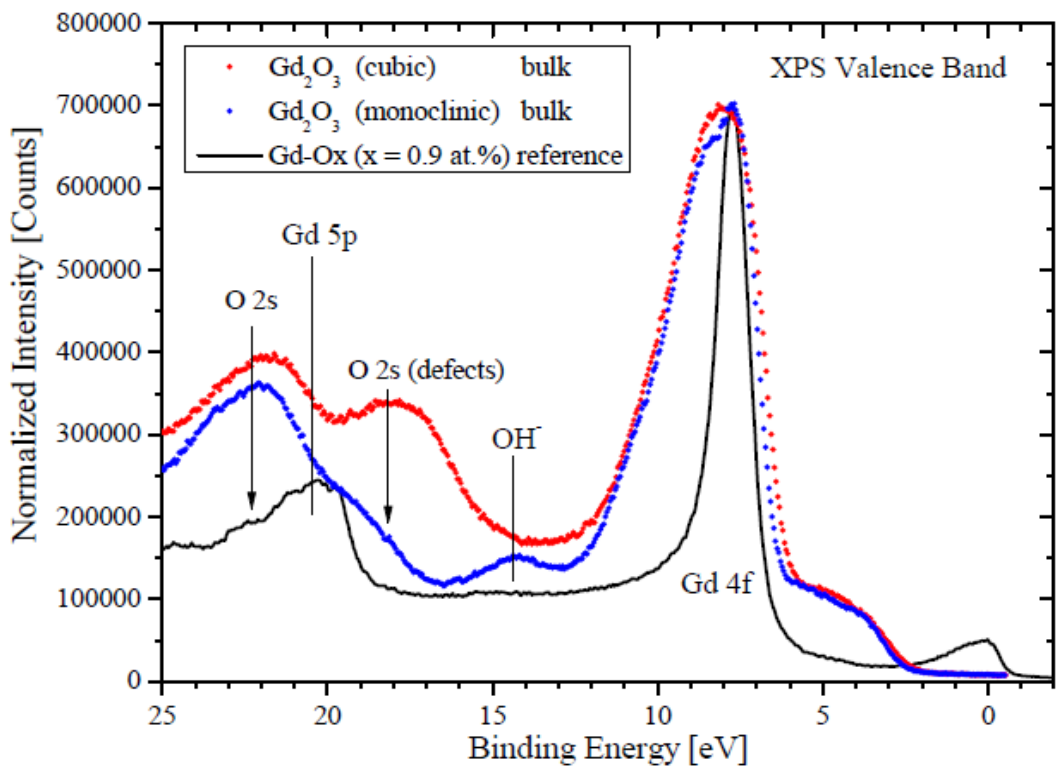

**Figure 7.** XPS valence bands (VBs) from c-$Gd_2O_3$ (cubic), m-$Gd_2O_3$ (monoclinic), and the Gd-$O_x$ (x = 0.9 at. %) XPS external standard.

The valence band (VB) area up to 25 eV was recorded for the samples and compared with that of the Gd-$O_x$ (x = 0.9 at. %) XPS external standard (see Fig. 7). The range of binding energies (BE) from 18 eV to 24 eV usually belongs to the O 2s electronic core-like states for the case of simple oxides [29-31,32]. However, in this part of the c- and m-$Gd_2O_3$ VB, the complex multiband overlapping can be seen, which is absent in Gd-O$_x$ (x = 0.9 at. %) BE region. Recently, Tang *et al*. [28] studied the peculiarities of the appearance of Gd 5p in the valence band of Gd@C$_{82}$(OH)$_x$, where x = 0; 12; and 22, and derived the Gd 5p contribution in the VB at ~23 eV. We detected the XPS band located at 22.51 eV, and it has relatively high intensity in the VBs of both c- and m-Gd$_2$O$_3$, while simultaneously being suppressed for Gd-O$_x$ by another band located at 20.92 eV (Fig. 7). Recall that the Gd-O$_x$ XPS external standard is essentially oxygen deficient, *i.e.*, it is nearly purely metallic. Therefore, the specificity of the band at 22.51 eV described above, as well as the XPS data presented in Refs. [29, 31-32], lead to the conclusion that this band represents O 2s states rather than Gd 5p, which are usually localized at 20.92 eV for metallic Gd (see Fig. 7, black spectrum). Oxidation of a metal usually results in a BE shift of metal electronic states to a higher BE region and transformation of the XPS line-shape in the most general cases of matter. Therefore, the band at 22.51 eV, which has the lowest intensity in Gd-O$_x$ (x = 0.9 at. %) and the highest intensities for c- and m-Gd$_2$O$_3$, is not only of "pure" O 2s origin because of the partial contribution from Gd 5p due to band overlap. The origin of the band at 22.51 eV might be understood using atomic calculations of photoionization cross-sections [40], according to which σ (Gd 5p) : σ (O 2s) = 1.1 : 0.019. which leads us to the conclusion regarding the dominating character of the contribution from the Gd 5p states in this VB region. On this basis, the band at 22.51 eV might be interpreted as the O 2s + Gd 5p contribution to the VB with a majority of Gd 5p states. At the same time, the band at 18.71 eV has dissimilar manifestations in the VBs of c- and m-Gd$_2$O$_3$ and is totally absent for Gd-O$_x$ (x = 0.9 at. %). We assume the origin of this band is oxygen defects because it has the same manifestation characteristics in the VB as the oxygen defect band at 532 eV in the O 1s core-level spectrum (see Figs. 6 and 7). Both of these bands exhibit a decrease in intensity in the spectra of m-Gd$_2$O$_3$ and have higher intensity for c-Gd$_2$O$_3$ leading us to assume that they share a common reason for their manifestation. In connection with this one can suppose that m-Gd$_2$O$_3$ with an excess of oxygen is likely has less defects and thus O 2s peak of c-Gd$_2$O$_3$ is higher as compared to m-

Gd$_2$O$_3$. The low-intensity and broad band at 14.32 eV is due to [Gd…O-OH$^-$] bonds because it is present solely in the XPS VB for monoclinic gadolinia, which was fabricated employing the "wet" technology process (see *Sample preparation …* details). The so-called OH$^-$ feature can be distinctly seen as well in the O 1s core-level spectrum of the same sample and is totally absent in c-Gd$_2$O$_3$, coinciding well with the interpretation offered above.

Finalizing our XPS VB mapping, we will now analyze the valence base-band area (BBA), which was essentially transformed in the gadolinium oxides compared to the nearly purely metallic Gd-O$_x$ (see Fig. 6, the BE range from 3 eV to 11 eV). Usually, this range belongs to hybridized O 2p (metal electronic states), and this hybridization might be considered one of the reasons for base-band width (BBW) broadening while transitioning from pure metal to its oxide form. This can be clearly seen in Fig. 6 in the mentioned BBA; an additional low-intensity and broad band is centered at ~6.8 eV (it is partially overlapped with a high-intensity peak), and the relatively narrow twin-peak band can be seen (the first sharp maximum is at 7.9 eV with an asymmetric shoulder at 8.5 eV, which broadened this band compared to that for nearly pure metallic gadolinium). The origin of the broad band with a spectral center at ~6.8 eV might be linked with only O 2p partial states, which contributed to the Gd–O bond so that they overlapped with the Gd 4f double peaks at 7.9 eV and 8.5 eV. Taking into account the results of the O 1s core-level and O 2s core-like states analyses (the derivation concerning oxygen defects and oxygen deficient clusters), the sharp Gd 4f peak at 7.9 eV coincides very well with that for nearly purely metallic Gd-O$_x$ (x = 0.9 at. %). At the same time, the manifestation of the strongly asymmetric shoulder at 8.5 eV (see Fig. 6) allows us to assume that this is also from the Gd 4f electronic states, but from the gadolinium oxide phase, where more or less complete oxidation occurs. In other words, we observe the contributions from both nearly metallic Gd 4f states because of the oxygen deficiency and Gd 4f from c-Gd$_2$O$_3$ and m-Gd$_2$O$_3$. The difference between the bands at 7.9 eV and 8.5 eV agrees well with the BE shift of metal electronic states to the higher BE region and the transformation of the XPS line-shape in the most general cases of matter. In addition, the essential asymmetry of this band might manifest not only because of the oxidation, but also because of the s-CK 4d – 4f4f decay channel ($N_{4,5} – N_{6,7}N_{6,7}$) of the excited electronic states [12] with regard to the joint OKT-van de Laan model [12, 34]. Finally, no contradictions among the XPS core-levels analysis and the XPS VB electronic structure mappings were found, and the XPS-based considerations agree well with the theoretical ones mentioned above.

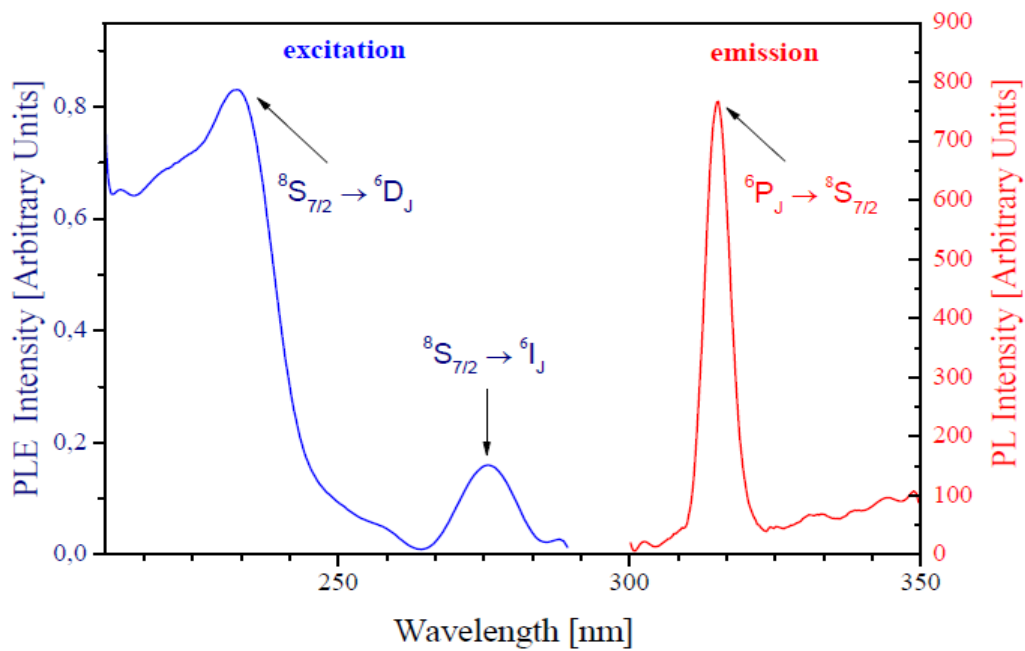

**Figure 8.** Photoluminescence excitation and emission spectra of c-$Gd_2O_3$ at room temperature.

*3.5. Luminescent Properties*

The luminescent properties of the c- and m-$Gd_2O_3$ polymorphs were studied with an optical spectroscopy technique. The cubic polymorph exhibits an intrinsic luminescence due to "defective" $Gd^{3+}$ ions at 315 nm. The luminescence is excited because of the 4f–4f intracentral mechanism at 232 nm and 275 nm (see Fig. 8). An existence of $Gd^{3+}$ local electronic levels in the so-called optical transparence area of $Gd_2O_3$ host-matrix is reasonably explained that these ions are essentially dissimilar with that for regular lattice cations. The energy scheme shown in Fig. 9 is explaining the excitation–luminescing relaxation of PL-mechanism in our current state of matter. We have to note that the luminescence of m-Gd$_2$O$_3$ polymorph is totally absent in the spectral range of 300–900 nm. Recall, that c- and m-Gd$_2$O$_3$ were fabricated using dissimilar technological synthesis methods (see *Sample preparation ...* details), so one should reasonably expect the dissimilar defectiveness degree as well in c- и m-Gd$_2$O$_3$ polymorphs. In other words, no luminescence excitation–relaxation processes of Gd$^{3+}$ ions were found in m-Gd$_2$O$_3$, allowing to derive that the cations are, probably, occupying mostly the regular structural positions in the lattice of m-Gd$_2$O$_3$ so having an undistorted electronic energy structure. This derivation well agrees with the XPS VB spectra of m- and c-polymorphs of gadolinia (see Fig. 7) according to which the m-Gd$_2$O$_3$ exhibits the formal electronic energy structure with essentially less defects (compare blue and red VB spectra at Fig. 7). As an alternative explanation the strong luminescence quenching might occur in m-polymorph because of the strong electrostatic interactions among electronic shells in Gd-ion as it had been theoretically predicted in Refs. [12, 33].

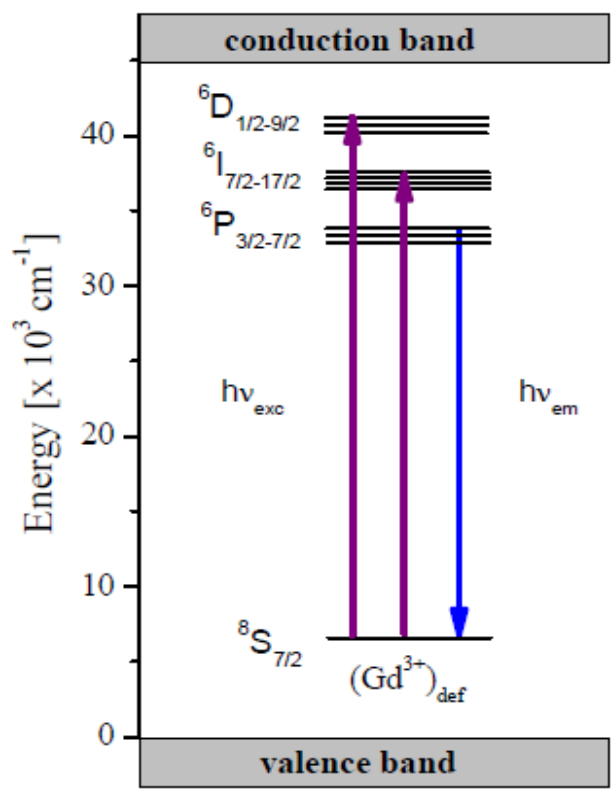

**Figure 9.** Luminescence Excitation – Relaxation scheme of PL-center in c-Gd$_2$O$_3$.

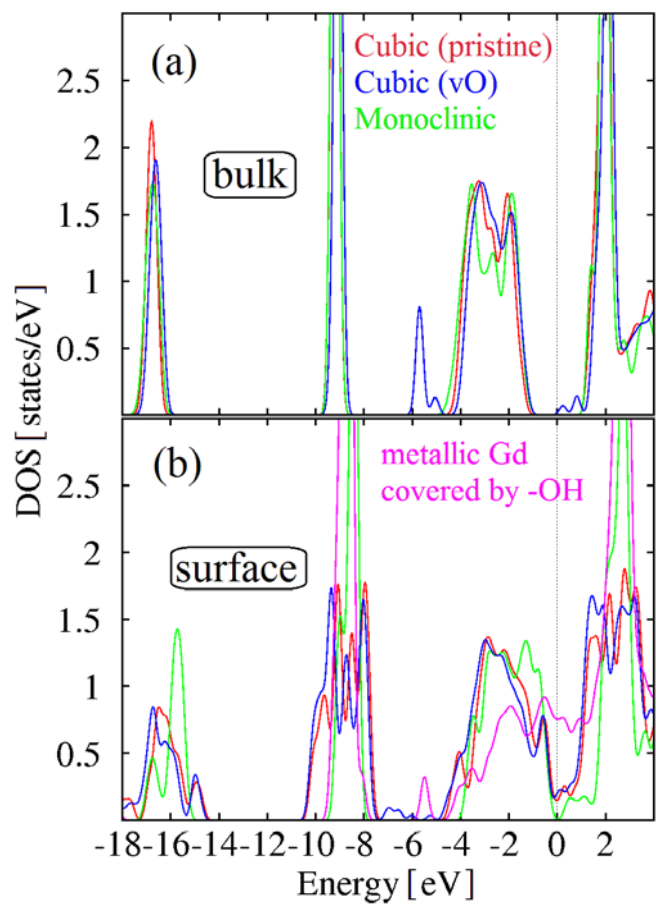

**Figure 10.** Total densities of states (DOS) for the bulk (a) and surface (b) configurations of the pristine cubic phase, monoclinic $Gd_2O_3$, and cubic $Gd_2O_3$ with oxygen vacancies (vO). The total DOS for the *surface* of metallic Gd passivated with hydroxyl […–OH] groups also are shown in panel (b).

*3.6. Electronic Structure Calculations*

The $Gd^{3+}$ local electronic levels exist in the so-called optical transparency area of the $Gd_2O_3$ host matrix because these ions are essentially dissimilar from regular lattice cations. The appearance of $Gd^{3+}$ levels in the band-gap of c-$Gd_2O_3$ might be due to strong distortions of the periodic distribution of the electronic DOS because of the localized point-defects of the crystalline lattice, such as oxygen vacancies in the surface states. This type of defect should be well-recognized in the O 2s DOS of the XPS valence band spectrum, and indeed, they are clearly seen in our results (see Fig. 7, the VB spectrum of c-$Gd_2O_3$) and in accordance with the results of theoretical modelling (see below).

At the same time, the photoluminescence of the $Gd^{3+}$ ion centers in the spectral range of 300–900 nm is absent in m-$Gd_2O_3$, probably because of the absence of specific oxygen defects. This assumption agrees well with the relatively weak XPS intensity of the O 2s band in the valence band spectrum of the m-polymorph. Another significant reason that might be considered for PL-quenching is the existence of nearly isolated [...O-OH$^-$] clusters. These clusters are also well-recognized by XPS because of the anomalous chemical bonding and corresponding XPS band in the O 1s core-level spectrum (see Fig. 6).

The data obtained with theoretical calculations lead to the conclusion that the presence of oxygen vacancies causes some electronic states to appear above and below the Fermi level (see Fig. 10a) in the case of the *bulk* electronic structure for the cubic and monoclinic $Gd_2O_3$ phases. The *surface* configuration (Fig. 10b) is characterized by the broadened peaks in both phases, which is in good agreement with experimental VB spectra (Fig. 7). These additional peaks are corresponding with violation of oxygen coordination of $Gd^{3+}$-ions. The opposite is seen for the case of the *bulk* electronic structure near the Fermi level in the cubic and monoclinic phases; the monoclinic phase secondary peak is now located at approximately 1 eV below the Fermi level, whereas the cubic phase is approximately 0.5 eV below. From the heights of these peaks, we can estimate the admixture bulk-to-surface ratio in both phases. In the experimental XPS spectra of the c- and m-phases, the intensity ratio of the main and secondary peaks in both phases is nearly 7:1. However, from theory, for the *surfaces*, this ratio should be 2:1 (cubic phase) and 1:1 (monoclinic phase). Therefore, for the cubic phase, the bulk-to-surface ratio will be 5:1, and hence, 6:1 for the monoclinic phase. Taking into account the slab thickness employed for the calculation of the *surface* (approximately 0.9 nm), we estimate and declare the size of the $Gd_2O_3$ nanoparticles to be 10-15 nm, which is in agreement with the experimental results. We cannot neglect the variant where a relatively small concentration of metallic gadolinium phase is fabricated as a result of the explosive pyrolysis process yielding monoclinic $Gd_2O_3$, so we performed the calculations for the (001) *surface* of metallic Gd, which was passivated with hydroxyl groups. These calculation results (see Fig. 10b) clearly demonstrate that the presence of the metallic phase provides at least two contributions to the final electronic structure. One contribution is the essential broadening of the valence bands toward the Fermi level, and the second one is the sharp peak at approximately -6 eV, which might be linked with hydroxyl groups (see discussion above).

Based on the comparison of the $Gd_2O_3$ electronic structures, we established that the transformation from bulk to nanoparticles provides an increased contribution from the *surface* and *subsurface* areas. This fits well with the formation of new electronic states in the valence and conduction bands near the Fermi level, which might be the reason for the photo-luminescing transitions. As it follows from the PL experiment, the transitions with wavelengths of 225 nm and 265 nm correspond to the energies of 3.03 eV and 2.56 eV, respectively, between the levels at the valence and conduction bands region. The luminescence emission at a wavelength of 315 nm corresponds well to an ~2.16 eV difference between energy levels. The source of these transitions is additional peaks related with the surface states (Fig. 10b) inside the gap between valence and conduction bands. Because of the lack of standard DFT-based methods to exact evaluation of the bandgap value we could only point appearance of additional energy states without further exact interpretation of the relation between electronic structure and optical properties. Thus, we can conclude that, for the $Gd_2O_3$ nanoparticles, new energy levels appearing in the vicinity of the Fermi level, which lead to an increased contribution from the *surface*, are the major reason for the radiative luminescence transitions, and the presence of oxygen vacancies established through experiments should provide an additional peak in the adsorption spectra.

## 4. Conclusions

Performing a combined analytic study using XRD, SEM, XPS, and PL leads to the following summary. Application of a synthesis method using standard precipitation from hydroxide gives single phase cubic $Gd_2O_3$ nanoparticles 50 nm in size with [Gd…O-O] defects that are well-recognized by XPS O 1s core-level and valence band measurements and PL spectroscopy. Based on the combined XPS-PL data obtained, an energy-band model of the intrinsic luminescence was suggested for c-$Gd_2O_3$, which is in good agreement with DFT calculations of the electronic structure and theoretical considerations regarding the origin of luminescence in the c-$Gd_2O_3$ polymorph. In contrast to c-$Gd_2O_3$, m-$Gd_2O_3$ deposited through explosive, pyrolysis essentially had a denser atomic structure assembled from three-dimensional complex nano-agglomerates without clear-edged boundaries. This produced m-$Gd_2O_3$ nanoparticles 21 nm in size having a cubic phase admixture in a concentration of only 2 at. % in the form of 15 nm primary edge-boundary particles. For the majority monoclinic $Gd_2O_3$ structure, we also obtained […O-OH⁻] clusters that were nearly isolated because of the "wet" technology applied during the first stage of explosive pyrolysis, and these defects are likely the reason for PL quenching in this sample. Both synthesized polymorphs demonstrated dissimilar contributions to the *charge-transfer* processes in comparison with the Gd-$O_x$ XPS standard because of the dissimilar origin of Gd-O chemical bonding and different types of defects in their atomic structures. The DFT-based model calculations performed here agreed well with the experimental data and supported the origin of PL in the c-$Gd_2O_3$ polymorph. Finally, all presented and discussed results fit well into the framework of the joint OKT-van der Laan model.

## Acknowledgments

The $Gd_2O_3$ sample synthesis, SEM, XRD, and PL measurements were made under support of the Ministry of Education and Science of the Russian Federation (Government Task No. 3.1485.2017/4.6). D.W.B. acknowledges support from the Ministry of Education and Science of the Russian Federation, Project № 3.7372.2017/БЧ. The XPS electronic structure qualifications were supported by the Russian Science Foundation (Project No. 14-22-00004). The authors gratefully acknowledge the technical support and scientific equipment provided by the Ural Center for Shared Use "Modern Nanotechnology" (SNSM Ural Federal University, Yekaterinburg, Russia).